\documentclass[12pt]{article}
\usepackage{amssymb}

\usepackage{graphicx}
\usepackage{amsmath}
\usepackage{amscd,amsthm}

\newtheorem{theorem}{Theorem}

\newtheorem{definition}[theorem]{Definition}

\newtheorem{lemma}[theorem]{Lemma}

\begin{document}

\title{Propagation of Waves in Networks of Thin Fibers}

\author{S. Molchanov, B. Vainberg \thanks{
The authors were supported partially by the NSF grant
DMS-0706928.} \and
Dept. of Mathematics, University of North Carolina at Charlotte, \and %
Charlotte, NC 28223, USA}

\date{}
\maketitle

\begin{abstract}
The paper contains a simplified and improved version of the
results obtained by the authors earlier. Wave propagation is
discussed in a network of branched thin wave guides when the
thickness vanishes and the wave guides shrink to a one dimensional
graph. It is shown that asymptotically one can describe the
propagating waves, the spectrum and the resolvent in terms of
solutions of ordinary differential equations on the limiting
graph. The vertices of the graph correspond to junctions of the
wave guides. In order to determine the solutions of the ODE on the
graph uniquely, one needs to know the gluing conditions (GC) on
the vertices of the graph.

Unlike other publications on this topic, we consider the situation
when the spectral parameter is greater than the threshold, i.e.,
the propagation of waves is possible in cylindrical parts of the
network. We show that the GC in this case can be expressed in
terms of the scattering matrices related to individual junctions.
The results are extended to the values of the spectral parameter
below the threshold and around it.
\end{abstract}

\section{Introduction}

Consider the stationary wave (Helmholtz) equation
\begin{equation}
H_{\varepsilon }u=-\varepsilon ^{2}\Delta u=\lambda u,\text{ \ \ \ }x\in
\Omega _{\varepsilon },\text{ \ \ \ }Bu=0\text{ \ \ on }\partial \Omega
_{\varepsilon }\text{,}  \label{h0}
\end{equation}
in a domain $\Omega _{\varepsilon }\subset R^{d},$ $d\geq 2,$ with
infinitely smooth boundary (for simplicity), which has the following
structure: $\Omega _{\varepsilon }$ is a union of a finite number of
cylinders $C_{j,\varepsilon }$ (which will be called channels) of lengths $%
l_{j},$ $1\leq j\leq N,$ with diameters of cross-sections of order $O\left(
\varepsilon \right) $ and domains $J_{1,\varepsilon },\cdots
,J_{M,\varepsilon }$ (which will be called junctions) connecting the channels
into a network. It is assumed that the junctions have diameters of the same
order $O(\varepsilon )$. The boundary condition has the form: $B=1$ (the
Dirichlet BC) or $B=\frac{\partial }{\partial n}$ (the Neumann BC) or $%
B=\varepsilon \frac{\partial }{\partial n}+\alpha (x),$ where $n$ is the
exterior normal and\ the function $\alpha \geq 0$ is real valued and does
not depend on the longitudinal (parallel to the axis) coordinate on the
boundary of the channels. One also can impose one type of BC on the lateral
boundary of $\Omega _{\varepsilon }$ and another BC on the free ends (which are
not adjacent to a junction) of the channels.

The axes of the
channels form edges $\Gamma _{j},$ $1\leq j\leq N,$ of the limiting $\left(
\varepsilon \rightarrow 0\right) $ metric graph $\Gamma $. The junctions
shrink to vertices of the graph $\Gamma $ when $\varepsilon \rightarrow 0.$
We denote the set of vertices $v_{j}$ by $V$.
Let $m$ channels have infinite length ($m=0$ for bounded $\Omega
_{\varepsilon }$). We start the numeration of $C_{j,\varepsilon }$ with the
infinite channels. So, $l_{j}=\infty $ for $1\leq j\leq m.$

\begin{figure}[tbph]
\begin{center}
\includegraphics[width=0.8\columnwidth]{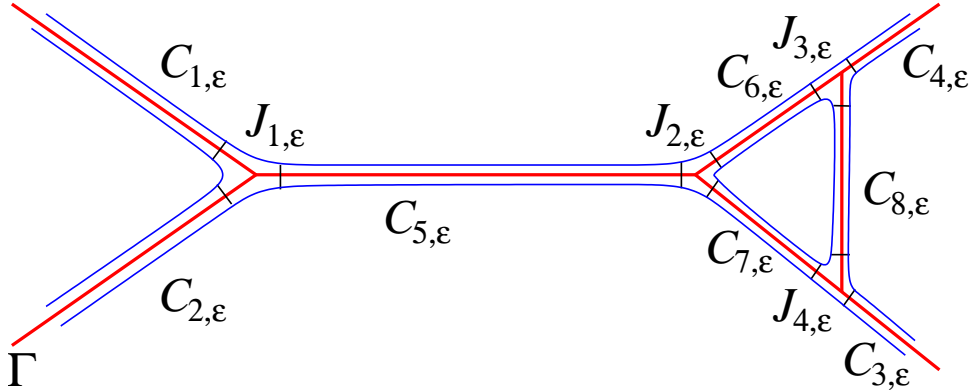}
\end{center}
\caption{An example of a domain $\Omega _{\protect\varepsilon }$ with four
junctions, four unbounded channels and four bounded channels.}
\label{fig-1}
\end{figure}

The goal of the paper is the asymptotic analysis of the spectrum
of $H_{\varepsilon },$ the resolvent $(H_{\varepsilon }-\lambda
)^{-1},$ and solutions of the corresponding non-stationary
problems for the
heat and wave equations as $\varepsilon \rightarrow 0.$ One can expect that $%
H_{\varepsilon }$ is close (in some sense) to a one dimensional operator on
the limiting graph $\Gamma $ with appropriate gluing conditions (GC) at the
vertices $v\in V.$ The ODE on $\Gamma $ appears in a natural way from the
following principle: the oscillating modes in the wave guides survive as $%
\varepsilon \rightarrow 0$ and the exponentially decaying and growing modes
disappear. However, the justification of this fact is not always simple. In
order to determine the solutions of ODE on $\Gamma $ uniquely, one needs to
know the GC on the vertices of $\Gamma $. The form of the GC in the general
situation was discovered quite recently in our papers \cite{MV}-\cite{MV2}.
It turned out that they can be expressed in terms of scattering matrices for problems of the wave
propagation through individual junctions of $\Omega _{\varepsilon }$. These
GC hold in all the cases: in the bulk of the spectrum $\lambda >\lambda _{0}$,
and near the threshold $\lambda \approx \lambda _{0}$, for bounded and
unbounded $\Omega _{\varepsilon }$.

Equation (\ref{h0}) degenerates when $\varepsilon =0.$ One could omit $%
\varepsilon ^{2}$ in (\ref{h0}). However, the problem under consideration
would remain singular, since the domain $\Omega _{\varepsilon }$ shrinks to
the graph $\Gamma $ as $\varepsilon \rightarrow 0.$ The presence of this
coefficient in the equation is convenient, since it makes the spectrum less
vulnerable to changes in $\varepsilon .$ As we shall see, in some important
cases (spider domains $\Omega _{\varepsilon }$) the spectrum of the problem (\ref{h0}) does not depend on $\varepsilon
,$ and the spectrum in the same cases will be magnified by a factor of $\varepsilon ^{-2}$ if $%
\varepsilon ^{2}$ in (\ref{h0}) is omitted. The operator $H_{\varepsilon
}=-\varepsilon ^{2}\Delta $ introduced in (\ref{h0}) will be considered as the operator in $L^{2}(\Omega _{\varepsilon })$.

An important class of domains $\Omega _{\varepsilon }$ are the self-similar
domains with only one junction and all the channels of infinite length. We
shall call them \textit{spider domains}. Thus, if $\Omega _{\varepsilon }$
is a spider domain, then there exists a point $\widehat{x}=x(\varepsilon )$
and an $\varepsilon $-independent domain $\Omega $ such that

\begin{equation}
\Omega _{\varepsilon }=\{(\widehat{x}+\varepsilon x):x\in \Omega \},
\label{hom}
\end{equation}
i.e. $\Omega _{\varepsilon }$ is the $\varepsilon $-contraction of $\Omega
=\Omega _{1}.$

For the sake of simplicity, we shall assume that $\Omega _{\varepsilon }$ is
self-similar in a neighborhood of each junction. Namely, let $%
J_{j(v),\varepsilon }$ be the junction which corresponds to a vertex $v\in V$
of the limiting graph $\Gamma .$ Consider a junction $J_{v,\varepsilon
}=J_{j(v),\varepsilon }$ and all the channels adjacent to $J_{v,\varepsilon
} $. If some of these channels have finite length, we extend them to
infinity. We assume that, for each $v\in V,$ the resulting domain $\Omega
_{v,\varepsilon },$ which consists of the junction $J_{v,\varepsilon }$ and
the semi-infinite channels emanating from it, is a spider domain. We also
assume here that all the channels $C_{j,\varepsilon }$ have the same
cross-section $\omega _{\varepsilon }.\ $This assumption is needed only to
make the results more transparent (The general case is studied in \cite{MV1}%
). From the self-similarity assumption it follows that $\omega _{\varepsilon
}\ $is an $\varepsilon -$homothety of a bounded domain $\omega \subset
R^{d-1}$. \

Let $\lambda _{0}<\lambda _{1}\leq \lambda _{2}...$ be eigenvalues of the
negative Laplacian $-\Delta _{d-1}$ in $\omega $ with the BC $B_{0}u=0$ on $%
\partial \omega $ where $B_{0}$ coincides with the boundary operator $B$ on
the channels, see (\ref{h0}), with $\varepsilon =1$ in the case of the third
boundary condition. Let $\{\varphi _{n}(y)\},$ $y\in \omega \in R^{d-1},$\
be the set of corresponding orthonormal eigenfunctions.\ Then $\lambda _{n}$
are eigenvalues of $-\varepsilon ^{2}\Delta _{d-1}$ in $\omega _{\varepsilon
}$ and $\{\varepsilon ^{-d/2}\varphi _{n}(y/\varepsilon )\}$ are the
corresponding eigenfunctions$.$ $\ $In the presence of infinite channels,
the spectrum of the operator $H_{\varepsilon }$ consists of an absolutely
continuous component which coincides with the semi-bounded interval $%
[\lambda _{0},\infty )$ and a discrete set of eigenvalues. The eigenvalues
can be located below $\lambda _{0}$ and can be embedded into the absolutely
continuous spectrum$.$ We will call the point $\lambda =\lambda _{0}$ the
threshold since it is the bottom of the absolutely continuous spectrum or
(and) the first point of accumulation of the eigenvalues as $\varepsilon
\rightarrow 0.$ Let us consider two of the simplest examples: the Dirichlet
problem in a half infinite cylinder and in a bounded cylinder of length $
l.$ In the first case, the spectrum of the negative Dirichlet Laplacian in $
\Omega _{\varepsilon }$\ is pure absolutely continuous and has multiplicity $
n+1$ on the interval $[\lambda _{n},\infty ).$ In the second case the
spectrum consists of the set of eigenvalues $\lambda _{n,m}=\lambda
_{n}+\varepsilon ^{2}m^{2}/l^{2},$ $n\geq 0,$ $m\geq 1.$

It was shown in \cite{MV}-\cite{MV1} that the wave propagation governed by
the operator $H_{\varepsilon },$ $\varepsilon \rightarrow 0$, as well as the
asymptotic behavior of the resolvent $(H_{\varepsilon }-\lambda )^{-1}$ and of the eigenvalues of
$H_{\varepsilon }$ above $\lambda_0$ can
be described in terms of the scattering solutions. While many particular
cases of that problem with $\lambda =\lambda _{0}+O(\varepsilon ^{2})$ or $
\lambda <\lambda _{0}$ were considered (see \cite{IT}-\cite{RS}), the
publications \cite{MV}-\cite{MV2} were the first ones dealing with the case $
\lambda \geq \lambda _{0},$ and the first ones where the significance of the
scattering solutions for asymptotic analysis of $H_{\varepsilon }$ was
established. In particular, it was shown there that in both cases $\lambda
>\lambda _{0}$ and $\lambda \thickapprox \lambda _{0},$ the GC of the
operator on the limiting graph $\Gamma $ will be expressed in terms of the
scattering matrices of the auxiliary problems on the spider domains
associated to individual junctions. A more profound analysis of the case $
\lambda \thickapprox \lambda _{0}$ can be found in \cite{MV2}.

The main goal of this paper is to overview the results of
\cite{MV1}-\cite{MV2} and simplify the proofs. We will mostly deal with the
case of $\lambda \in (\lambda _{0},\lambda _{1})$ where the results and
proofs are more transparent. The number of scattering solutions is the
smallest in this case and the scattering matrix is of the smallest size
(compared to the case $\lambda >\lambda _{1}$). One of our main results is as follows.
\begin{theorem}
 If $\lambda _{0} \leq \lambda \leq
\lambda _{1}$, then the resolvent $(H^{(1)}_{\varepsilon }-\lambda)^{-1}$ can be approximated by $(H^{(1)}_{\varepsilon }-(\lambda - \lambda_0))^{-1},$ where $H^{(1)}_{\varepsilon}=-\varepsilon ^{2}\frac{d^{2}}{dt^{2}}$ is the operator of the second derivative defined on functions $\varsigma$ on the limiting graph $\Gamma$ with the GC of the form
\[
i\varepsilon \lbrack I_{v}+T_{v}(\lambda )]\frac{d}{dt}\varsigma^v(0)-%
\sqrt{\lambda -\lambda _{0}}[I_{v}-T_{v}(\lambda )]\varsigma^v (0)=0,
\]
Here $T_{v}(\lambda )$ is the scattering matrix of the axillary problem on the spider domain which corresponds to the junction $J_{v,\varepsilon}$, and $\varsigma^v$ is the vector which consists of restrictions of the function $\varsigma$ (defined on $\Gamma$) onto edges adjacent to $v$.

To be more exact, for any compactly supported $f$, the following relation is valid on channels outside of the support of $f$ with
exponential accuracy
\[
(H_{\varepsilon }-\lambda)^{-1}f \sim [(H^{(1)}_{\varepsilon }-
(\lambda-\lambda_0))^{-1}f_0]\varphi _{0}(y/\varepsilon ),
~~\varepsilon \rightarrow 0,~~f_0=<f,\varphi _{0}(y/\varepsilon )>.
\]
\end{theorem}
A more accurate statement of this theorem as well as some of its generalizations will be given in section 5.

Note that the eigenvalues of the problem in $\Omega _{\varepsilon }$ are
located not only below the threshold, but also above it. They depend on $%
\varepsilon $ and move very fast on the $\lambda $-axis as $\varepsilon
\rightarrow 0.$ Thus one can not expect to obtain an asymptotic
approximation of the resolvent $(H_{\varepsilon }-\lambda )^{-1}$ when $%
\lambda =\lambda ^{\prime }>\lambda _{0}$ is fixed and $\varepsilon
\rightarrow 0.$ An asymptotic approximation of the resolvent $%
(H_{\varepsilon }-\lambda )^{-1}$ as $\varepsilon \rightarrow 0$ can be
valid only if an exponentially small (in $\varepsilon $), but depending on $%
\varepsilon$, set on the $\lambda $-axis is omitted. Another option is to
fix $\lambda =\lambda ^{\prime }>\lambda _{0}$ and pass to the limit as $%
\varepsilon \rightarrow 0$ without $\varepsilon $ taking values in some small
set which depends on $\lambda ^{\prime }$.

While the condition\ $\lambda >\lambda _{0}$ is natural for the wave
propagation, the properties of the heat and diffusion processes depend on
spectrum of $H_{\varepsilon }$ near\ $\lambda =\lambda _{0}.$ As a by-product of the simpler approach to the problem introduced below, we will get
a better result concerning the asymptotic behavior of the eigenvalues of $%
H_{\varepsilon }$ in bounded domains $\Omega _{\varepsilon }$\ as $%
\varepsilon \rightarrow 0,$\ $\lambda =\lambda _{0}+O(\varepsilon ^{2}).$ It
was shown in \cite{MV1}, \cite{MV2} that the main terms of the eigenvalues
of $H_{\varepsilon }$ when\ $\lambda =\lambda _{0}+O(\varepsilon ^{2}),$ $%
\varepsilon \rightarrow 0,$ coincide with the eigenvalues of the
operator on the limiting graph with the GC defined by the
scattering matrix at $\lambda =\lambda _{0}$. An explicit
description of GC at $\lambda =\lambda _{0}$ for arbitrary
junctions (of order $O(\varepsilon)$) was also given there.
Significantly later (see publications in arXiv), some of our results where repeated in \cite{Gr}.
The new elements there are the location of the eigenvalues below the
threshold and more accurate asymptotics of eigenvalues near the threshold. We
will show here that the approach used in
\cite{MV1} and \cite{MV2} provides an approximation of the
eigenvalues near the threshold with an exponential accuracy as well as
the location of the eigenvalues below the threshold.

The plan of the paper is the following. The elliptic problem in
$\Omega_{\varepsilon }$ with a fixed $\varepsilon =1$ will be
studied in the next section. In particular, the scattering
solutions are defined there. The asymptotic behavior of the
resolvent $(H_{\varepsilon }-\lambda )^{-1}$, of the spectrum and of the
scattering solutions as $\varepsilon \rightarrow 0,$\ $\lambda
>\lambda _{0},$ are obtained in section 3 for the simplest domains with one junction (spider domains).
The one dimensional problem on the limiting graph will be studied
in section 4. The case of arbitrary domains $\Omega_{\varepsilon
}$ is considered in section 5. The last section is devoted
studying the spectrum near the threshold.

\section{Scattering solutions and analytic properties of the resolvent when $%
\protect\varepsilon $ is fixed.}

We introduce Euclidean coordinates $(t,y)$ in channels $C_{j,\varepsilon }$
chosen in such a way that the $t$-axis is parallel to the axis of the
channel (so, $t$ is not time, but space variable!), hyperplane $R_{y}^{d-1}$
is orthogonal to the axis, and $C_{j,\varepsilon }$ has the following form
in the new coordinates:
\begin{equation*}
C_{j,\varepsilon }=\{(t,\varepsilon y):0<t<l_{j},\text{ }y\in \omega \}.
\end{equation*}
If a channel $C_{j,\varepsilon }$ is bounded ($l_{j}<\infty $), the
direction of the $t$ axis can be chosen arbitrarily (at least for now). If a channel is unbounded, then $t=0$ corresponds to its
cross-section which is attached to the junction.

Let us recall the definition of scattering solutions for the problem (\ref
{h0}) in $\Omega _{\varepsilon }$ when $\lambda \in (\lambda _{0},\lambda
_{1}).$ Consider the non-homogeneous problem
\begin{equation}
(-\varepsilon ^{2}\Delta -\lambda )u=f,\text{ \ }x\in \Omega _{\varepsilon };%
\text{ \ \ \ }Bu=0\text{ \ on }\partial \Omega _{\varepsilon }.  \label{a1}
\end{equation}

\begin{definition}
\label{d1}Let $f\in L_{com}^{2}(\Omega _{\varepsilon })$ have a compact
support, and $\lambda <\lambda _{1}$.\ A solution $u$ of (\ref{a1}) is
called outgoing if it has the following asymptotic behavior at infinity in
each infinite channel $C_{j,\varepsilon },$ \ \ $1\leq j\leq m$:
\begin{equation}
u=a_{j}e^{i\frac{\sqrt{\lambda -\lambda _{0}}}{\varepsilon }t}\varphi
_{0}(y/\varepsilon )+O(e^{-\frac{\alpha t}{\varepsilon }}),\text{ \ \ }%
\alpha >0.  \label{a2}
\end{equation}
\end{definition}

\noindent \textbf{Remarks.} 1. Here and everywhere below we assume that
\begin{equation}
\text{Im}\sqrt{\lambda -\lambda _{0}}\geq 0.  \label{imp}
\end{equation}
Thus, outgoing solutions decay at infinity if $\lambda <\lambda _{0}$.

2. Obviously, if (\ref{a2}) holds with some $\alpha >0,$ then it holds with
any $\alpha <\sqrt{\lambda _{1}-\lambda }.$

\begin{definition}
\label{d2}Let $\lambda <\lambda _{1}.$ A function $\Psi =\Psi
_{p}^{(\varepsilon )},$ $1\leq p\leq m,$ is called a solution of the
scattering problem in $\Omega _{\varepsilon }$ if
\begin{equation}
(-\varepsilon ^{2}\Delta -\lambda )\Psi =0,\text{ \ }x\in \Omega
_{\varepsilon };\text{ \ \ \ }B\Psi =0\text{ \ on }\partial \Omega
_{\varepsilon },  \label{b9}
\end{equation}
and $\Psi $ has the following asymptotic behavior in infinite channels $%
C_{j,\varepsilon },\ \ 1\leq j\leq m:$%
\begin{equation}
\Psi _{p}^{(\varepsilon )}=[\delta _{p,j}e^{-i\frac{\sqrt{\lambda -\lambda
_{0}}}{\varepsilon }t}+t_{p,j}e^{i\frac{\sqrt{\lambda -\lambda _{0}}}{%
\varepsilon }t}]\varphi _{0}(y/\varepsilon )+O(e^{-\frac{\alpha t}{%
\varepsilon }}),~~t\rightarrow \infty ,\text{ \ }\alpha >0.  \label{b10}
\end{equation}
Here $\delta _{p,j}$ is the Kronecker symbol, i.e. $\delta _{p,j}=1$ if $%
p=j, $ $\delta _{p,j}=0$ if $p$ $\neq j.$
\end{definition}
\noindent \textbf{Remark.} If $\lambda _{0}<\lambda <\lambda _{1},$ then the term with
the coefficient $\delta _{p,j}$ in (\ref{b10}) corresponds to the incident
wave (coming through the channel $C_{p,\varepsilon }$), the terms with
coefficients $t_{p,j}$ describe the transmitted waves. The transmission
coefficients $t_{p,j}=t_{p,j}(\varepsilon ,\lambda )$ depend on $\varepsilon
$ and $\lambda $. The matrix

\begin{equation}
T=[t_{p,j}]  \label{scm}
\end{equation}
is called\textit{\ the scattering matrix}. Note that the scattering solution
and scattering matrix are defined for all $\lambda <\lambda _{1}.$ We
assume that Im$\sqrt{\lambda -\lambda _{0}}>0$\ when $\lambda <\lambda _{0},$
and the incident wave is growing (exponentially) at infinity in this case.

The outgoing and scattering solutions are defined similarly when $\lambda
\in (\lambda _{n},\lambda _{n+1})$ (see \cite{MV1})$.$ In this case, any
outgoing solution has $n+1$ waves in each channel propagating to infinity
with the frequencies $\sqrt{\lambda -\lambda _{s}}/\varepsilon ,$ $0\leq
s\leq n$. There are $m(n+1)$ scattering solutions: the incident wave may
come through one of $m$ infinite channels with one of $(n+1)$ possible
frequencies. The scattering matrix has the size $m(n+1)\times m(n+1)$ in
this case.

Standard arguments based on the Green formula provide the following
statement.

\begin{theorem}
\label{t3}When $\lambda _{0}<\lambda <\lambda _{1},$\ the scattering matrix $%
T$ is unitary and symmetric ($t_{p,j}=t_{j,p}$).
\end{theorem}

The operator $H_{\varepsilon }=-\varepsilon ^{2}\Delta $, which corresponds
to the eigenvalue problem (\ref{h0}), is non-negative, and therefore the
resolvent
\begin{equation}
R_{\lambda }=(H_{\varepsilon }-\lambda )^{-1}:L^{2}(\Omega _{\varepsilon
})\rightarrow L^{2}(\Omega _{\varepsilon })  \label{res}
\end{equation}
is analytic in the complex $\lambda $-plane outside the positive semi-axis $%
\lambda \geq 0.$ If $\Omega _{\varepsilon }$ is bounded (all the channels
are finite), then operator $R_{\lambda }$ is meromorphic in $\lambda $ with
a discrete set $\Lambda =\{\mu _{j,\varepsilon }\}$ of poles of first order
at the eigenvalues $\lambda =\mu _{j,\varepsilon }$ of operator $%
H_{\varepsilon }.$ If $\Omega _{\varepsilon }$ has at least one infinite
channel, then the spectrum of $H_{\varepsilon }$ has more complicated
structure (see Theorem \ref{t1} below). In this case, the spectrum has an absolutely
continuous component $[\lambda _{0},\infty ),$ and the resolvent $R_{\lambda
}$ is meromorphic in $\lambda \in C\backslash \lbrack \lambda _{0},\infty ).$
We are going to consider the analytic extension of the operator $R_{\lambda
} $ to the absolutely continuous spectrum. One can extend $R_{\lambda }$
analytically from above (Im$\lambda >0$) or below, if it is considered as an
operator in the following spaces (with a smaller domain and a larger range):
\begin{equation}
R_{\lambda }:L_{com}^{2}(\Omega _{\varepsilon })\rightarrow
L_{loc}^{2}(\Omega _{\varepsilon }).  \label{b2}
\end{equation}
These extensions do not coincide on $[\lambda _{0},\infty )$. To be
specific, we always will consider extensions from the upper half plane (Im$%
\lambda >0$). We will call (\ref{b2}) truncated resolvent of the operator $%
H_{\varepsilon },$ since it can be identified with the resolvent (\ref{res})
multiplied from the left and right by a cut-off function.

\begin{theorem}
\label{t1} Let $\Omega _{\varepsilon }$ have at least one infinite channel.
Then

(1) The spectrum of the operator $H_{\varepsilon }$ consists of the
absolutely continuous component $[\lambda _{0},\infty )$ and, possibly, a
discrete set $\{\mu _{j,\varepsilon }\}$ of non-negative eigenvalues $%
\lambda =\mu _{j,\varepsilon }\geq 0$ with the only possible limiting point
at infinity.\ The multiplicity of the a.c. spectrum changes at points $%
\lambda =\lambda _{n},$ and is equal to $m(n+1)$ on the interval $(\lambda
_{n},\lambda _{n+1})$.

(2) The operator (\ref{b2}) admits a meromorphic extension from the upper
half plane Im$\lambda >0$ onto $[\lambda _{0},\infty )$ with the branch
points at $\lambda =\lambda _{n}$ of the second order and poles of first
order at $\lambda =\mu _{j,\varepsilon }.$ In particular, if $\lambda
_{n}\in \{\mu _{j,\varepsilon }\},$ then operator (\ref{b2}) has the form
\begin{equation*}
R_{\lambda }=\frac{A(n)}{\lambda -\lambda _{n}}+O(\frac{1}{\sqrt{|\lambda
-\lambda _{n}|}}),\ \lambda \rightarrow \lambda _{n}.
\end{equation*}

(3) If $f\in L_{com}^{2}(\Omega _{\varepsilon }),$ $\lambda <\lambda _{1},$
and $\lambda $ is not a pole or the branch point ($\lambda =\lambda _{0}$)
of \ the operator (\ref{b2}), then the problem (\ref{a1}), (\ref{a2}) is
uniquely solvable and the outgoing solution $u$ can be found as the $%
L_{loc}^{2}(\Omega _{\varepsilon })$ limit
\begin{equation}
u=R_{\lambda +i0}f.  \label{b3}
\end{equation}

(4) There exist exactly $m\ $ different scattering solutions for the values
of $\lambda <\lambda _{1}$ which are not a pole or the branch point of \ the
operator (\ref{b2}), and the scattering solution is defined uniquely after
the incident wave is chosen.

(5) The scattering matrix $T$ is analytic in $\lambda ,$ when $\lambda
<\lambda _{1},$ with a branch point of second order at $\lambda =\lambda
_{0} $ and without real poles.

The matrix $T$ is orthogonal if $\lambda <\lambda _{0}.$
\end{theorem}
\noindent \textbf{Remark.} Let $\lambda _{n}\notin \{\mu _{j,\varepsilon }\}.$ If the
homogeneous problem (\ref{a1}) with $\lambda =\lambda _{n}$ has a nontrivial
solution $u$ such that
\begin{equation}
u=a_{j}\varphi _{n}(y/\varepsilon )+O(e^{-\gamma t}),\text{ \ \ \ }x\in
C_{j,\varepsilon },\text{ \ \ }t\rightarrow \infty ,\text{ \ \ }1\leq j\leq
m, ~ \gamma >0,  \label{inf}
\end{equation}
then $R_{\lambda +i0}=\frac{B(n)}{\sqrt{\lambda -\lambda _{n}}}+O(1),$ \ $%
\lambda \rightarrow \lambda _{n}.$ If such a solution $u$ does not exist,
then operator (\ref{b2}) is bounded in a neighborhood of $\lambda =\lambda
_{n}.$

\noindent \textbf{Proof of Theorem \ref{t1}. }All the statements above concern the problem
with a fixed value of $\varepsilon $ and can be proved using standard
elliptic theory arguments. A detailed proof can be found in \cite{MV1}, a
shorter version is given below.

In order to prove the part of statement (1) concerning the absolutely
continuous spectrum of the operator $H=-\Delta $, we split the domain $%
\Omega _{\varepsilon }$ into pieces by introducing cuts along the bases $t=0$
of all infinite channels. We denote the new (not connected) domain by $%
\Omega _{\varepsilon }^{\prime },$ and denote the negative Dirichlet
Laplacian in $\Omega _{\varepsilon }^{\prime }$ by $H_{\varepsilon }^{\prime
},$ i. e. $H_{\varepsilon }^{\prime }$ is obtained from $H_{\varepsilon }$
by introducing additional Dirichlet boundary conditions on the cuts.
Obviously, the operator $H_{\varepsilon }^{\prime }$ has the absolutely
continuous spectrum described in statement (1) of the theorem. Since the
wave operators for the couple $H_{\varepsilon },$ $H_{\varepsilon }^{\prime }
$ exist and are complete (see \cite{BIR}), the operator $H_{\varepsilon }$ has
the same absolutely continuous spectrum as $H_{\varepsilon }^{\prime }.$

The remaining part of statement (1) and statements (2) and (3) can be proved
by one of the well known equivalent approaches based on a reduction of the
boundary problem (\ref{a1}) to a Fredholm equation which depends
analytically on $\lambda .$ For this purpose one can use a parametrix
(almost inverse operator): equation (\ref{a1}) is solved separately in
channels and junctions, and then the parametrix can be constructed from
those local inverse operators using a partition of unity (allowing one to
glue the local inverse operators), see \cite{MV1}. A similar approach is
based on gluing together these local inverse operators using
Dirichlet-to-Neumann maps on the cuts of the channels which were introduced
in the previous paragraph.

Statements (4) and (5) follow immediately from statement (3) and Theorem
\ref{t3}. Indeed, one can look for the solution $\Psi _{p}^{(\varepsilon )}$
of the scattering problem in the form
\begin{equation}
\Psi _{p}^{(\varepsilon )}=\chi e^{-i\frac{\sqrt{\lambda -\lambda _{0}}}{%
\varepsilon }t}\varphi _{0}(y/\varepsilon )+u  \label{fu}
\end{equation}
where $\chi \in C^{\infty }(\Omega _{\varepsilon }),$ $\chi =0$\ outside $%
C_{p,\varepsilon },$ $\chi =1$\ in $C_{p,\varepsilon }\cap \{t>1\}$. This
reduces problem (\ref{b9}), (\ref{b10}) to problem (\ref{a1}), (\ref{a2})
for $u$ with $f$\ supported on $C_{p,\varepsilon }\cap \{0\leq t\leq 1\}$.
Statement (3) of the theorem, applied to the latter problem, justifies
statement (4). Function $u,$ defined in (\ref{fu}), satisfies the
homogeneous equation (\ref{a1}) in infinite channels $C_{j,\varepsilon },$ $%
j\neq p,$ and in $C_{p,\varepsilon }\cap \{t>1\},$ and it is meromorphic at
the bottoms of these channels (at $t=0$ for $j\neq p,$ and $t=1$ when $j=p).$
Solving the problems in these channels by separation of variables, we obtain
that the scattering matrix $T$ is meromorphic in $\lambda ,$ when $\lambda
<\lambda _{1}$ with a branch point of second order at $\lambda =\lambda _{0}$%
. It also follows from here that $T$ is real valued when $\lambda <\lambda
_{0}.$ Analyticity of $T$ and Theorem \ref{t3} imply that $T$ is orthogonal
when $\lambda <\lambda _{0}.$ From the orthogonality ($\lambda <\lambda _{0}$%
) and unitarity ($\lambda _{0}<\lambda <\lambda _{1}$) of $T$ it follows
that $T$ does not have poles.\qed

\section{Spider domains, $\protect\varepsilon \rightarrow 0.$}

Let us recall that $\Omega _{\varepsilon }$ is called a spider
domain if it is self-similar (see (\ref{hom})) and consists of one
junction and several semi-infinite channels.

\begin{theorem}
\label{tsp}Let $\Omega _{\varepsilon }$ be a spider domain and $\lambda
<\lambda _{1}$. Then

(1) the eigenvalues $\lambda =\mu _{j,\varepsilon }=\mu _{j} $ of operator $%
H_{\varepsilon }$ and the scattering matrix $T$ do not depend on $%
\varepsilon $,

(2) the truncated resolvent (\ref{b2}) has the following estimate: if $f$ is
supported on $\varepsilon $-neighborhood of the junction, then
\begin{equation}
|R_{\lambda }f|\leq C\delta ^{-1}\varepsilon ^{-d/2}||f||_{L^{2}},~~\lambda
<\lambda _{1},\text{ \ \ }\delta =\text{dist}(\lambda ,\{\mu _{j}\}),
\label{res1}
\end{equation}
outside of $2\varepsilon $-neighborhood of the junction,

(3) the scattering solutions have the following form on the channels of the
domain:
\begin{equation}
\Psi _{p}^{(\varepsilon )}=[\delta _{p,j}e^{-i\frac{\sqrt{\lambda -\lambda
_{0}}}{\varepsilon }t}+t_{p,j}e^{i\frac{\sqrt{\lambda -\lambda _{0}}}{%
\varepsilon }t}]\varphi _{0}(y/\varepsilon )+r_{p,j}^{\varepsilon },\text{ \
\ \ }x\in C_{j,\varepsilon },\ \ 1\leq j\leq m,  \label{sce0}
\end{equation}
where $|r_{p,j}^{\varepsilon }|$ $\leq C\delta ^{-1}e^{-\alpha \frac{t}{%
\varepsilon }}$ when $\varepsilon >0,$ $\frac{t}{\varepsilon }\geq 1$ and $%
0\leq \lambda <\lambda _{1}.$ Here $\alpha <\sqrt{\lambda _{1}-\lambda }$ is
arbitrary, $C=C(\alpha )$.
\end{theorem}
\noindent \textbf{Remark.} Formula (\ref{sce0}) looks similar to the definition (\ref
{b10}). In fact, the remainder in (\ref{b10}) decays only when $t\rightarrow
\infty ,$ and (\ref{b10}) does not allow us to single out the main term of
asymptotics of scattering solutions as $\varepsilon \rightarrow 0.$

\noindent \textbf{Proof.} All the statements above follow immediately from
self-similarity of the domain $\Omega _{\varepsilon }$. Namely, we make the
substitution
\begin{equation}
x\rightarrow \frac{x-\widehat{x}}{\varepsilon }  \label{sub}
\end{equation}
(see (\ref{hom})) and reduce problem (\ref{a1}) in $\Omega _{\varepsilon }$
to the problem in $\Omega $ which corresponds to $\varepsilon =1.$ These two
problems have the same eigenvalues and scattering matrices. This
justifies the first statement. Let $v_{\lambda },$ $g$ be functions $%
R_{\lambda }f,f$ after the change of variables (\ref{sub}). From statement
(2) of Theorem \ref{t1} it follows that
\begin{equation*}
||v_{\lambda }||_{L^{2}(K)}\leq C\delta ^{-1}||g||_{L^{2}}=C\delta
^{-1}\varepsilon ^{-d/2}||f||_{L^{2}},
\end{equation*}
where $K$ consists of the parts of the channels of $\Omega $ where $1<t<3.$
Then the standard a priori estimates for the solutions of the equation $%
\Delta u-\lambda u=0$ imply the same estimate for $|v_{\lambda }|$ on the
cross sections $t=2:$
\begin{equation*}
|v_{\lambda }|\leq C\delta ^{-1}\varepsilon ^{-d/2}||f||_{L^{2}},\text{ \ \ }%
t=2.
\end{equation*}
The latter implies the same estimate for $|v_{\lambda }|$ when\ $t>2$ by
solving the equation $\Delta u-\lambda u=0$ in the corresponding regions of
the channels of $\Omega $ with the boundary condition at\ $t=2.$ This
justifies the second statement of Theorem \ref{tsp}. The last statement
can be proved absolutely similarly. We reduce the scattering problem in $%
\Omega _{\varepsilon }$ to the scattering problem in $\Omega $ and use
representation (\ref{fu}) with $\varepsilon =1.$ This implies (\ref{b10})
with $\varepsilon =1$ and the remainder term $r_{p,j}$ such that $|r_{p,j}|$
$\leq C\delta ^{-1}e^{-\alpha t}$ for $t>1.$\ It remains only to make the
substitution inverse to (\ref{sub}).\qed

In spite of its simplicity, Theorem \ref{tsp} allows us to obtain two very
important results: small $\varepsilon $ asymptotics of the spectrum of $%
H_{\varepsilon }$ and the resolvent $(H_{\varepsilon }-\lambda )^{-1}$ for
arbitrary networks of thin wave guides $\Omega _{\varepsilon }$. For this
purpose, we need to rewrite (\ref{sce0}) in a slightly different (less
explicit) form.

We denote by $\varsigma _{p,j}$ the linear combination of exponents in the
square brackets in (\ref{sce0}). This is a function on the edge $\Gamma _{j}$
of the graph. Let $\varsigma _{p}$ be the vector-column with components $%
\varsigma _{p,j},$ $1\leq j\leq m.$ Obviously, $\varsigma _{p}$ satisfies
the equation \
\begin{equation}
(\varepsilon ^{2}\frac{d^{2}}{dt^{2}}+\lambda -\lambda _{0})\varsigma =0.
\label{greq}
\end{equation}
We will use notation $\Psi _{p}^{(\varepsilon )}$ for both the scattering
solution and the column-vector whose components $\Psi _{p,j}^{(\varepsilon
)} $ are restrictions of the scattering solution $\Psi _{p}^{(\varepsilon )}$
on the channels $C_{j,\varepsilon },\ \ 1\leq j\leq m.$\ Then (\ref{sce0})
can be rewritten in the vector form as
\begin{equation}
\Psi _{p}^{(\varepsilon )}=\varsigma _{p}\varphi _{0}(y/\varepsilon
)+r_{p}^{(\varepsilon )}=[e_{p}e^{-i\frac{\sqrt{\lambda -\lambda _{0}}}{%
\varepsilon }t}+t_{p}e^{i\frac{\sqrt{\lambda -\lambda _{0}}}{\varepsilon }%
t}]\varphi _{0}(y/\varepsilon )+r_{p}^{(\varepsilon )},
\label{sce1}
\end{equation}
where $x\in
\cup _{1\leq j\leq m}C_{j,\varepsilon },~~r_{p}^{(\varepsilon )}$ is the vector with components $%
r_{p,j}^{(\varepsilon )},$ all components $e_{p,j}$ of the vector $e_{p}$
are zeroes except $e_{p,p}$ which is equal to one, and $t_{p}$ is the $p$-th
column of the scattering matrix $T.$\ Let us construct the $m\times m$
matrix with columns $\Psi _{p}^{(\varepsilon )}$ and the matrix $\varsigma $
with columns $\varsigma _{p},$ $1\leq p\leq m.$ As it is easy to see, $\varsigma
(0)=(I+T),$ $\varsigma ^{\prime }(0)=i\frac{\sqrt{\lambda -\lambda _{0}}}{%
\varepsilon }(-I+T),$ and therefore,
\begin{equation}
i\varepsilon (I+T)\varsigma ^{\prime }(0)-\sqrt{\lambda -\lambda _{0}}%
(I-T)\varsigma (0)=0.  \label{gc10}
\end{equation}
Of course, this equality also holds for individual columns $\varsigma _{p}$
of matrix $\varsigma .$

It is essential for extending the results to arbitrary networks of
wave guides $\Omega _{\varepsilon }$ that the gluing condition (\ref{gc10})
together with some condition at infinity is equivalent to the explicit form
of $\varsigma _{p}$ given by (\ref{sce1}). Namely, let $\varsigma $
satisfy (\ref{greq}). Then
\begin{equation*}
\varsigma =\alpha _{p}e^{-i\frac{\sqrt{\lambda -\lambda _{0}}}{\varepsilon }%
t}+\beta _{p}e^{i\frac{\sqrt{\lambda -\lambda _{0}}}{\varepsilon }t}
\end{equation*}
with some constant vectors $\alpha _{p},$ $\beta _{p}.$ We will say that $%
\varsigma =\psi _{p}$ is \textit{a solution of the scattering problem on the
graph $\Gamma $} with the incident wave coming through the edge $\Gamma _{p}$
if $\psi _{p}$ satisfies equation (\ref{greq}), GC (\ref{gc10}), and $\alpha
_{p}=e_{p},$ i.e.,
\begin{equation}
\psi _{p}=e_{p}e^{-i\frac{\sqrt{\lambda -\lambda _{0}}}{\varepsilon }%
t}+\beta _{p}e^{i\frac{\sqrt{\lambda -\lambda _{0}}}{\varepsilon }t}
\label{2222}
\end{equation}
Thus, we specify the incident wave and impose the GC defined by the
scattering problem in $\Omega _{\varepsilon }$, but we do not specify the
scattering coefficients of the outgoing wave. The next theorem shows that
the scattering problem on the graph will have the same scattering
coefficients as the problem on $\Omega _{\varepsilon }$.

\begin{theorem}
\label{t6} Formulas (\ref{sce0}), (\ref{sce1}) and $\Psi _{p}^{(\varepsilon
)}=\psi _{p}\varphi _{0}(y/\varepsilon )+r_{p}^{(\varepsilon )}$ are
equivalent.
\end{theorem}
\noindent \textbf{Proof}. It was already shown that $\varsigma _{p}$ defined in (\ref
{sce1}) satisfies (\ref{gc10}). Conversely, if we write $\beta _{p}$ in (\ref
{2222}) as $t_{p}+h_{p}$ and put (\ref{2222}) into (\ref{gc10}), we will get
that $h_{p}=0,$ i.e. $\psi _{p}$ coincides with $\varsigma _{p}$ defined in (%
\ref{sce1}). \qed

\section{One-dimensional problem on the graph.}

The spectrum of the operator $H_{\varepsilon }$ an the asymptotic behavior of
the resolvent will be expressed in terms of the solutions of a problem on
the limiting graph $\Gamma $ which is studied in this section.

Let $\Omega _{\varepsilon }$ be an arbitrary (bounded or unbounded) domain
described in the introduction, and let $\Gamma $ be the corresponding
limiting graph. Points of $\Gamma $ will be denoted by $\gamma $ with $t$\
being a parameter on each edge $\Gamma _{j}$ of the graph. We are going to
introduce a special spectral problem
\begin{equation}
h_{\varepsilon }\varsigma :=-\varepsilon ^{2}\frac{d^{2}}{dt^{2}}\varsigma
=(\lambda -\lambda _{0})\varsigma   \label{spp}
\end{equation}
on smooth functions $\varsigma =\varsigma (\gamma )$ on $\Gamma $ which
satisfy the following GC\ at vertices. We split the set $V$ of vertices $v$
of the graph into two subsets $V=V_{1}\cup V_{2},$ where the vertices from
the set $V_{1}$ have degree $1$ and correspond to the free ends of the
channels, and the vertices from the set $V_{2}$ have degree at least two and
correspond to the junctions $J_{v,\varepsilon }$. We keep the same BC at $%
v\in V_{1}$ as at the free end of the corresponding channel of $\Omega
_{\varepsilon }$ (see (\ref{h0})):
\begin{equation}
B\varsigma =0\text{ \ \ at }v\in V_{1}.  \label{bce}
\end{equation}

The GC at each vertex $v\in $ $V_{2}$ will be defined in terms of an
auxiliary scattering problem for a spider domain $\Omega _{v,\varepsilon
}^{\prime }.$ This domain is formed by the individual junction $%
J_{v,\varepsilon }$ which corresponds to the vertex $v,$ and all channels
with an end at this junction, where the channels are extended to infinity if
they have a finite length. Let $T=T_{v}(\lambda )$ be the scattering matrix
for the problem (\ref{h0}) in the spider domain $\Omega _{v,\varepsilon
}^{\prime }$ and let $I_{v}$ be the unit matrix of the same size as the size
of $T.$ We choose the parametrization on $\Gamma $ in such a way that $t=0$
at $v$ for all edges adjacent to this particular vertex$.$ Let $d=d(v)\geq 2$%
\ be the order (the number of adjacent edges) of the vertex $v\in V_{2}.$
For any function $\varsigma $ on $\Gamma ,$ we form a column-vector $%
\varsigma ^{(v)}=\varsigma ^{(v)}(t)$ with $d(v)$\ components which is
formed by the restrictions of $\varsigma $ on the edges of $\Gamma $
adjacent to $v.$ We will need this vector only for small values of $t\geq
0. $ The components of the vector $\varsigma ^{(v)}$ are taken in the same
order as the order of channels of $\Omega _{v,\varepsilon }^{\prime }.$ The
GC at the vertex $v\in V_{2}$ has the form
\begin{equation}
i\varepsilon \lbrack I_{v}+T_{v}(\lambda )]\frac{d}{dt}\varsigma ^{(v)}(t)-%
\sqrt{\lambda -\lambda _{0}}[I_{v}-T_{v}(\lambda )]\varsigma ^{(v)}(t)=0,%
\text{ \ \ \ }t=0,\text{ \ \ \ }v\in V_{2},  \label{gc}
\end{equation}
if $\lambda \neq \lambda _{0}.$ Condition (\ref{gc}) can degenerate if $%
\lambda =\lambda _{0},$ and it requires some regularization in this case.

Solutions of (\ref{spp}) have the following form
\begin{equation*}
\varsigma =a_{j}e^{i\frac{\sqrt{\lambda -\lambda _{0}}}{\varepsilon }%
t}+b_{j}e^{-i\frac{\sqrt{\lambda -\lambda _{0}}}{\varepsilon }t},~\gamma \in
\Gamma _{j}.
\end{equation*}
If Im$\lambda >0$ and $\varsigma \in L^{2}(\Gamma ),$ then $b_{j}=0$ for
infinite edges (see (\ref{imp})). Thus, if $\varsigma $ satisfies
equation (\ref{spp}) in a neighborhood of infinity, then
\begin{equation}
\varsigma =a_{j}e^{i\frac{\sqrt{\lambda -\lambda _{0}}}{\varepsilon }t},%
\text{ \ }\gamma \in \Gamma _{j},\text{ \ \ }1\leq j\leq m,~~t>>1.
\label{bes}
\end{equation}
We will assume that condition (\ref{bes}) holds also when $\lambda $ is
real, i.e., we consider only those solutions of (\ref{spp}) with real $%
\lambda =\lambda ^{\prime }>\lambda _{0}$ which can be obtained as the limit
of solutions with complex $\lambda =\lambda ^{\prime }+i\varepsilon $ when $%
\varepsilon \rightarrow 0$.

We will call function $g=g_{\lambda }(\gamma ,\xi ;\varepsilon ),$ $\gamma
,\xi \in \Gamma ,$ \textit{the Green function of the problem (\ref{spp})-(%
\ref{bes})} if it satisfies the equation (with respect to variable $\gamma ):
$%
\begin{equation}
-\varepsilon ^{2}\frac{d^{2}}{dt^{2}}g-(\lambda -\lambda _{0})g=\delta _{\xi
}(\gamma ),  \label{stl}
\end{equation}
and conditions (\ref{bce})-(\ref{bes}). Here $\xi $ is a point of $\Gamma $
which is not a vertex, and $\delta _{\xi }(\gamma )$ is the delta function
supported on $\gamma =\xi .$

\begin{lemma}
\label{lsym}Let $\lambda <\lambda _{1},$ $\lambda \neq \lambda _{0}.$
\bigskip Operator $h_{\varepsilon }=-\varepsilon ^{2}\frac{d^{2}}{dt^{2}}$
is symmetric on the space of smooth, compactly supported functions on $%
\Gamma $ which satisfy conditions (\ref{bce}) and (\ref{gc}).
\end{lemma}
\noindent \textbf{Proof.} One needs only to show that
\begin{equation}
\left\langle \frac{d}{dt}\varsigma _{1}^{(v)}(t),\varsigma
_{2}^{(v)}(t)\right\rangle -\left\langle \varsigma _{1}^{(v)}(t),\frac{d}{dt}%
\varsigma _{2}^{(v)}(t)\right\rangle =0,\text{ \ \ \ }t=0,\text{ \ \ \ }v\in
V_{2},  \label{symm}
\end{equation}
for any two vector functions $\varsigma =\varsigma _{1}^{(v)},$ $\varsigma
=\varsigma _{1}^{(v)}$ which satisfy GC\ (\ref{gc}) (similar relation at $%
v\in V_{1}$ obviously holds). Let $\lambda \in (\lambda _{0},\lambda _{1}).$
Then matrix $T_{v}(\lambda )$ is unitary (Theorem \ref{t3}). If matrix $%
I_{v}+T_{v}$ is non-degenerate, we rewrite\ (\ref{gc}) in the form $\frac{d}{%
dt}\varsigma ^{(v)}(t)=A\varsigma ^{(v)}(t),$ $t=0,$ where the matrix
\begin{equation*}
A=\frac{\sqrt{\lambda -\lambda _{0}}}{i\varepsilon }[I_{v}+T_{v}(\lambda )%
]^{-1}[I_{v}-T_{v}(\lambda )]
\end{equation*}
is real. The latter immediately implies (\ref{symm}). Similar arguments can
be used if $I_{v}-T_{v}$ is non-degenerate. If both matrices are degenerate
(i.e., $T_{v}$ has both eigenvalues, $\pm 1$), we consider a unitary matrix $%
U $ such that $UT_{v}U^{\ast }$ is a diagonal unitary matrix. Since $%
\left\langle U\varsigma _{1},U\varsigma _{2}\right\rangle =\left\langle
\varsigma _{1},\varsigma _{2}\right\rangle $ for any two vectors $\varsigma
_{1},\varsigma _{2},$ one can easily reduce the proof of (\ref{symm}) to the
case when $T_{v}$ is a diagonal unitary matrix. Then \ (\ref{gc}) implies
the following relations for \ coordinates $\varsigma _{j}(t)$ of the vector $%
\varsigma ^{(v)}(t):$ $\varsigma _{j}^{\prime }(0)=a_{j}\varsigma _{j}(0)$
or $\varsigma _{j}(0)=b_{j}\varsigma _{j}^{\prime }(0)$, where constants $%
a_{j},b_{j}$ are real. The first case occurs if the corresponding diagonal
element of $T_{v}$ differs from $-1,$ the second relation is valid if this
element is $-1.$ These relations for $\varsigma _{j}(t)$ imply (\ref{symm}).
Similar arguments can be used to prove (\ref{symm}) when $\lambda <\lambda
_{0},$ since matrix $T_{v}$ is orthogonal in this case (see Theorem \ref{t1}%
). \qed

\begin{theorem}
\label{tr}For any $\varepsilon >0$ there is a discrete set $\Lambda
(\varepsilon )$ on the interval $[-\lambda _{0},\lambda _{1})$ such that the
Green function $g_{\lambda }(\gamma ,\xi ;\varepsilon )$ exists for all $%
\lambda <\lambda _{1},$ $\lambda \notin \Lambda (\varepsilon ),$ and has the
form
\begin{equation}
g_{\lambda }=\frac{h(\gamma ,\xi ,\lambda ,\varepsilon )}{D(\lambda
,\varepsilon )},  \label{qw}
\end{equation}
where function $h$ is continuous on the set $\gamma ,\xi \in \Gamma
,~\lambda <\lambda _{1},~\varepsilon >0$ and uniformly bounded on each
bounded subset, and
\begin{equation}
D(\lambda ,\varepsilon )=\Sigma _{m=1}^{N}c_{m}(\lambda )e^{i\frac{\sqrt{%
\lambda -\lambda _{0}}}{\varepsilon }s_{m}}.  \label{fd}
\end{equation}
Here $s_{m}$ are constants, functions $c_{m}(\lambda )$ are analytic in $%
\lambda <\lambda _{1}$ with a branch point of second order at $\lambda
=\lambda _{0},$ and $D\neq 0$\ if $\lambda <\lambda _{0}.$
\end{theorem}
\noindent \textbf{Proof.} We fix the parametrization on each edge $\Gamma _{j}$ of the
graph. Then, obviously,

\begin{equation}
g_{\lambda }=a_{j}e^{-i\frac{\sqrt{\lambda -\lambda _{0}}}{\varepsilon }%
t}+b_{j}e^{i\frac{\sqrt{\lambda -\lambda _{0}}}{\varepsilon }t},\text{ \ }%
\gamma \in \Gamma _{j},~~~\text{if}~~\xi \notin \Gamma _{j},  \label{rl1}
\end{equation}
\begin{equation}
g_{\lambda }=a_{j}e^{-i\frac{\sqrt{\lambda -\lambda _{0}}}{\varepsilon }%
t}+b_{j}e^{i\frac{\sqrt{\lambda -\lambda _{0}}}{\varepsilon }t}+\frac{%
\varepsilon }{\sqrt{\lambda -\lambda _{0}}}\sin [\frac{\sqrt{\lambda
-\lambda _{0}}}{\varepsilon }(t-\tau )_{-}],~~~\text{if}~~\xi \in \Gamma
_{j}.  \label{rl2}
\end{equation}
Here $\tau $ is the coordinate of the point $\xi ,~(t-\tau )_{-}=\min
(t-\tau ,0)$, and the last term in (\ref{rl2}) is a particular solution of (%
\ref{stl}) on $\Gamma _{j}$ with a bounded support. There are $2N$ unknown
constants in the formulas above where $N$\ is the total number of edges of
the graph. Conditions (\ref{bce})-(\ref{bes}) provide $2N$ linear equations
for these constants. As it is easy to see, the coefficients for unknowns in all
the equations have the form $a(\lambda )e^{i\frac{\sqrt{\lambda -\lambda _{0}%
}}{\varepsilon }s}$, where $a(\lambda )$ is analytic in $\lambda <\lambda _{1}
$ with a branch point of second order at $\lambda =\lambda _{0},$ and $s=0$
or $\pm l_{j}$ ($l_{j}$ are the lengths of the finite channels). The
exponential factors in the coefficients appear when the formulas (\ref{rl1}%
), (\ref{rl2}) are substituted into GC at the end point of the edge $\Gamma
_{j}$ where $t=l_{j}.$ We apply Cramer's rule to solve this system of $2N$
equations. This immediately provides all the statements of the theorem with $%
D(\lambda ,\varepsilon )$ being the determinant of the system. One only
needs to show that $D\neq 0$\ for $\lambda <\lambda _{0}.$ Note that the
latter fact implies the discreteness of the set $\Lambda (\varepsilon
)=\{\lambda :D(\lambda ,\varepsilon )=0\}.$

Obviously, $D=0$ if and only if the homogeneous problem (\ref{spp})-(\ref
{bes}) has a non-trivial solution. Let $\lambda <\lambda _{0}.$ Then
solutions $\varsigma $\ of the problem (\ref{spp})-(\ref{bes}) decay at
infinity, and
\begin{eqnarray}
0 &=&\int_{\Gamma }[-\varepsilon ^{2}\varsigma ^{\prime \prime }-(\lambda
-\lambda _{0})\varsigma ]\varsigma d\gamma   \notag \\
&=&-\varepsilon ^{2}\Sigma _{v}\left\langle \frac{d}{dt}\varsigma
^{(v)},\varsigma ^{(v)}\right\rangle |_{v}+\int_{\Gamma }[\varepsilon
^{2}(\varsigma ^{\prime })^{2}-(\lambda -\lambda _{0})\varsigma ^{2}]d\gamma
.  \label{dnez}
\end{eqnarray}

It was shown in the proof of Lemma \ref{lsym} that it is enough to consider
only diagonal matrices $T$ when the terms under the sign $\Sigma _{v}$ above
with $v\in V_{2}$ are evaluated. Since $T$ is orthogonal when $\lambda
<\lambda _{0},$ the diagonal elements of $T$ are equal to $\pm 1.$ Then (\ref
{gc}) means that each component of the vector $\varsigma ^{(v)}$\ or its
derivative is zero at the vertex. Hence, the terms in the sum above with $%
v\in V_{2}$ are equal to zero. They are zeroes also for those $v\in V_{1}$
where the boundary condition in (\ref{bce}) is the Dirichlet or Neumann
condition. If $v\in V_{1}$ and $B=\varepsilon \frac{d}{dt}+a,~a\geq 0,$
these terms are non-positive. Hence, relation (\ref{dnez}) implies that $%
\varsigma =0$ when $\lambda <\lambda _{0}$. \qed

Theorem \ref{tr} does not contain a statement concerning the structure of
the discrete set $\Lambda (\varepsilon ).$ This set becomes more and more
dense when $\varepsilon \rightarrow 0.$ In general, every point $\lambda
^{\prime }\in $($\lambda _{0},\lambda _{1}$)\ belongs to $\Lambda
(\varepsilon )$ for some sequence of$\ \varepsilon =\varepsilon _{j}(\lambda
^{\prime })\rightarrow 0.$ However, it is not an absolutely arbitrary discrete
set, but the set of zeroes of a specific analytic function (\ref{fd}), and
this fact provides the following restriction on the set $\Lambda
(\varepsilon ).$

\begin{lemma}
\label{ld}$\ $\ For each bounded interval $[\alpha ,\lambda _{1}]$, each $%
\sigma >0$ and some $M,$ there are $c\varepsilon ^{-1}$ intervals $I_{j}$ of
length $\sigma $ such that
\begin{equation*}
|D(\lambda ,\varepsilon )|>c\sigma ^{M}\text{ \ \ when \ }\varepsilon >0,%
\text{ \ }\lambda \in \lbrack \alpha ,\lambda _{1}]\backslash \cup I_{j},%
\text{ \ }c=c(\alpha )\text{.}
\end{equation*}
\end{lemma}

This Lemma is a particular case of Lemma 15 from \cite{MV1} (the set $\Gamma
_{0}$ is empty in the case which is considered in this paper).

In order to construct the resolvent of the problem in $\Omega _{\varepsilon
},$ we need to represent the Green function $g_{\lambda }$ of the problem (%
\ref{stl}), (\ref{bce})-(\ref{bes}) on the graph $\Gamma $ through the
solutions of the scattering problems on the spider subgraphs of $\Gamma .$

We will call a function $\psi =\psi _{p}(\gamma )$ \textit{solution of the
scattering problem on the graph $\Gamma $} if it satisfies the equation (\ref
{spp}), conditions (\ref{bce}), (\ref{gc}) and has the following form at
unbounded edges of the graph:
\begin{equation}
\psi _{p}(\gamma )=\delta _{p,j}e^{-i\frac{\sqrt{\lambda -\lambda _{0}}}{%
\varepsilon }t}+a_{p,j}e^{i\frac{\sqrt{\lambda -\lambda _{0}}}{\varepsilon }%
t},\text{ \ }\gamma \in \Gamma _{j},\text{ \ \ }1\leq p,j\leq m,  \label{scg}
\end{equation}
where $\delta _{p,j}$ is the Kronecker symbol. This scattering solution
corresponds to the wave coming through the edge $\Gamma _{p}.$ These
scattering solutions on the graphs were introduced in the previous section
in the case when the graph corresponds to a spider domain. In fact, only
this simple case will be needed below.

\begin{lemma}
\label{lw}If the graph $\Gamma $ corresponds to a spider domain $\Omega
_{\varepsilon },$ then the scattering solution $\psi _{p}(\gamma )$ exists
and is defined uniquely for all $\lambda <\lambda _{1},$ $\lambda \neq
\lambda _{0}.$ Any function $\varsigma $ on $\Gamma $ which satisfies
equation (\ref{spp}) and GC condition (\ref{gc}) is a linear combination of
the scattering solutions $\psi _{p}(\gamma ).$
\end{lemma}
\noindent \textbf{Remark.} For arbitrary graphs, one may have nontrivial solutions of
the homogeneous problem (\ref{spp})-(\ref{bes}) supported on the set of
bounded edges of the graph. This occurs when $\lambda \in \Lambda
(\varepsilon ).$ The set $\Lambda (\varepsilon )$ is empty for spider graphs.
\textbf{Proof.} If we take $a_{p,j}=t_{p,j},$ where $t_{p,j}$ are the
scattering coefficients in the spider domain $\Omega _{\varepsilon },$ then
function (\ref{scg}) will satisfy (\ref{gc}) (see the derivation of (\ref
{gc10})). Hence, the scattering solutions $\psi _{p}(\gamma )$ exist for all
$\lambda <\lambda _{1},$ $\lambda \neq \lambda _{0},$ since the scattering
coefficients are defined for those $\lambda $ by Theorem \ref{t1}. If we put
function (\ref{scg}) with $a_{p,j}=t_{p,j}+h_{p,j}$ into GC \ (\ref{gc}), we
immediately get that $h_{p,j}=0$ (see the proof of Theorem \ref{t6}). Thus,
scattering solutions are defined uniquely. The space of solutions of
equation (\ref{spp}) is $2m$-dimensional. The $m\times 2m$ dimensional
matrix ($I_{v}+T_{v}(\lambda ),$ $I_{v}+T_{v}(\lambda )$) formed from
coefficients in GC\ (\ref{gc}) has rank $m.$ Hence, the solution space of
the problem (\ref{spp}), (\ref{gc}) is $m$-dimensional. Obviously, functions
$\psi _{p}$ are linearly independent on $\Gamma .$ Thus any solutions of (%
\ref{spp}), (\ref{gc}) is a linear combination of functions $\psi _{p}.$
\qed

Let $\Gamma _{j_{0}}$ be the edge of $\Gamma $ which contains the point $\xi
$ (see (\ref{stl})). We cut the graph $\Gamma $ into simple graphs $\Gamma
(v)$ with one vertex $v$ by cutting all the bounded edges at some points $%
\xi _{j}\in \Gamma _{j}.$ We will choose $\xi _{j_{0}}=\xi .$ Let us denote
by $\Gamma^{\prime }(v)$ the spider graph which is obtained by extending all
the edges of $\Gamma (v)$ to infinity. Let $\psi _{p,v}(\gamma )$ be the
scattering solutions on the graph $\Gamma ^{\prime }(v).$

\begin{lemma}
\label{lgr}There exist functions
\begin{equation*}
a=a_{p,v}(\lambda ,\varepsilon ,\xi ),\text{ \ }\lambda <\lambda _{1},\text{
}\varepsilon >0,\text{ }\xi \in \Gamma _{j_{0}},
\end{equation*}
which are continuous, bounded on each bounded set, and such that
\begin{equation*}
g_{\lambda }=\Sigma _{p}\frac{a_{p,v}(\lambda ,\varepsilon ,\xi )}{D(\lambda
,\varepsilon )}\psi _{p,v}(\gamma ),\text{ \ \ }\gamma \in \Gamma (v).
\end{equation*}
\end{lemma}
\noindent \textbf{Proof}. It follows from the previous lemma that $g_{\lambda }$ can
be represented as a linear combination of the scattering solutions:
\begin{equation*}
g_{\lambda }=\Sigma _{p}c_{p,v}\psi _{p,v}(\gamma ),\ \ \gamma \in \Gamma
(v).
\end{equation*}
In order to find the coefficients $c_{p,v}$, we note that $g_{\lambda }$ is
equal to a combination of two exponents on the edge $\Gamma _{p}\subset
\Gamma (v)$ with the coefficient of the incident wave equal to $c_{p,v}:$%
\begin{equation*}
g_{\lambda }=c_{p,v}e^{i\frac{\sqrt{\lambda -\lambda _{0}}}{\varepsilon }%
t}+b_{p,v}e^{-i\frac{\sqrt{\lambda -\lambda _{0}}}{\varepsilon }t},\text{ \ }%
\gamma \in \Gamma _{p}\subset \Gamma (v).
\end{equation*}
Now $c_{p,v}$ can be found by comparing the formula above and (\ref{qw}) at
two points of $\Gamma _{p}.$ \qed

\section{Small $\protect\varepsilon $ asymptotics for the problem in $\Omega
_{\protect\varepsilon }$.}

As everywhere above, the domain $\Omega _{\varepsilon },$ considered below,
can be bounded or unbounded. Denote by $\Lambda ^{0}$ the union of
eigenvalues of the operator (\ref{a1}) in all the spider domains $\Omega
_{v,\varepsilon }^{\prime }$ associated to $\Omega _{\varepsilon }.$ These
spider domains consist of individual junctions and all the channels adjacent
to this junction. The channels are extended to infinity if they have a
finite length. The set $\Lambda ^{0}$ does not depend on $\varepsilon $ due
to Theorem \ref{tsp}. Let us recall that $\Lambda (\varepsilon )$ is the set
of eigenvalues of the one dimensional problem (\ref{spp})-(\ref{bes}) on the
limiting graph (see Theorem \ref{tr}).

The eigenvalues of the operator $H_{\varepsilon }=-\varepsilon ^{2}\Delta $
of the problem (\ref{h0}) which are located on the interval $(-\infty
,\lambda _{1})$ are exponentially close to the set $\Lambda ^{0}\cup \Lambda
(\varepsilon ).$ In the process of proving this statement, we will get the
asymptotic approximation of the resolvent $(H_{\varepsilon }-\lambda )^{-1}$
as $\varepsilon \rightarrow 0.$\ Namely, the following theorem will be
proved.

Let $\lambda ^{\prime }<\lambda _{1}$ and let $\Lambda ^{\nu }$ be $e^{\frac{%
-\nu \alpha }{\varepsilon }}$-neighborhood of the set $\Lambda ^{0}\cup
\Lambda (\varepsilon ).$ Assume that the right-hand side of (\ref{a1}) has a
compact support which is separated from junctions, i.e. there exist $\tau
,d>0$ such that support of $f$ belongs to $\cup \Delta _{j}$ where $\Delta
_{j}$ is the part of the channel $C_{j,\varepsilon }$ defined by the
inequalities $\tau \leq t\leq l_{j}-\tau $ if $l_{j}<\infty ,$ or $\tau \leq
t\leq d$ if the channel is infinite.

\begin{theorem}
\label{tR1}(1) There exists $\nu >0$ such that the eigenvalues $\mu
_{j,\varepsilon }$ of the operator $H_{\varepsilon }$ which belong to the
interval $(-\infty ,\lambda ^{\prime })$ with an arbitrary $\lambda ^{\prime
}<\lambda _{1}$ are located in $e^{\frac{-\nu \alpha }{\varepsilon }}$%
-neighborhood of the set $\Lambda ^{0}\cup \Lambda (\varepsilon ).$ Here $%
\alpha =\lambda _{1}-\lambda ^{\prime }.$

(2) Let the support of $f$ belong to $\cup \Delta _{j}$ and let $u=R_{\lambda }f$
be the solution of problem (\ref{a1}). Here $R_{\lambda }$ is the truncated
resolvent (\ref{res}). Then for any $\eta >0,$ there exist $\nu >0$ and $%
\rho =\rho (\eta )>0$ such that $u=R_{\lambda }f\ $\ has the following
asymptotic behavior in all the channels outside $\eta -$neighborhood of the
support of $f$
\begin{equation}
u=R_{\lambda }f=(\widehat{g}_{\lambda }f_{0})\varphi _{0}(\frac{y}{%
\varepsilon })+O(e^{\frac{-\rho }{\varepsilon }}),\text{ \ \ \ }\lambda \in
(-\infty ,\lambda ^{\prime })\backslash \Lambda ^{\nu },\text{ \ \ }%
\varepsilon \rightarrow 0.  \label{rem}
\end{equation}
Here
\begin{equation*}
f_{0}=f_{0}(\gamma )=\left\langle f,\varepsilon ^{-d/2}\varphi _{0}(\frac{y}{%
\varepsilon })\right\rangle ,\text{ \ \ }\gamma \in \Gamma ,
\end{equation*}
and $\widehat{g}_{\lambda }$ is the integral operator on the graph $\Gamma $
whose kernel is the Green function $g_{\lambda }$ constructed in Theorem
\ref{tr}:
\begin{equation*}
\widehat{g}_{\lambda }f_{0}=\int_{\Gamma }g_{\lambda }(\gamma ,\xi
;\varepsilon )f_{0}(\xi )d\xi .
\end{equation*}
\end{theorem}
\noindent \textbf{Remark.}  Below, we also will get the asymptotics of $u=R_{\lambda }f$
on the support of $f$, as well as a more precise estimate of the remainder
in (\ref{rem}).

\noindent \textbf{Proof. }Let
\begin{equation*}
f_{1}=f_{1}(x)=f-\varepsilon ^{-d/2}f_{0}\varphi _{0}(\frac{y}{\varepsilon }%
),\text{ \ \ }x\in \Omega _{\varepsilon },
\end{equation*}
i.e. $f_{0}=f_{0}(\gamma )$ is the first Fourier coefficient of the
expansion of $f$ with respect to the basis $\{\varepsilon ^{-d/2}\varphi
_{j}(\frac{y}{\varepsilon })\},$ and $f_{1}$ is the sum of all the terms of
the expansion without the first one. We are going to show that $u=R_{\lambda
}f$ has the following form on the channels of $\Omega _{\varepsilon }:$
\begin{equation}
u=R_{\lambda }f=(\widehat{g}_{\lambda }f_{0})\varphi _{0}(\frac{y}{%
\varepsilon })+\chi R_{\lambda }^{0}f_{1}+O(e^{\frac{-\rho }{\varepsilon }%
}),\ \ \ \lambda \in (-\infty ,\lambda ^{\prime })\backslash \Lambda ^{\nu
},\ \ \varepsilon \rightarrow 0,  \label{lll}
\end{equation}
where $\nu ,\rho >0,$ $\chi \in C^{\infty }(\Omega _{\varepsilon })$ is a
cut off function such that $\chi =0$ on all the junctions, $\chi =1$ outside
of $\varepsilon $-neighborhood of junctions, and \ function $R_{\lambda
}^{0}f_{1}$ is defined by solving the following simple problem in the infinite cylinder.
Let $f_{1,j}$ be the restriction of $f_{1}$ onto the channel $%
C_{j,\varepsilon }.$ We extend \ the channel $C_{j,\varepsilon }$ to
infinity (in both directions) and extend $f_{1,j}$ by zero. Let $u_{j}$ be
the outgoing solution of the equation
\begin{equation*}
-\varepsilon ^{2}\Delta u-\lambda u=f_{1,j}
\end{equation*}
in the extended channel. Then $R_{\lambda }^{0}f_{1}$ is
defined as $R_{\lambda }^{0}f_{1}=u_{j}$ in the channel $
C_{j,\varepsilon }$ Obviously,
$\chi R_{\lambda }^{0}f_{1}$ can be considered as a function on $\Omega
_{\varepsilon }.$

The justification of (\ref{lll}) and the proof of the Theorem \ref{tR1} are
based on an appropriate choice of the parametrix (''almost inverse
operator''):
\[
P_{\lambda }:L_{\tau ,d}^{2}\rightarrow L_{loc}^{2}(\Omega
_{\varepsilon }),
\]
which is defined as follows
\begin{equation}
P_{\lambda }f=(\widehat{G}_{\lambda }f_{0})\varphi _{0}(\frac{y}{\varepsilon
})+(\chi R_{\lambda }^{0}f_{1})-\Sigma _{v}\chi _{v}R_{\lambda ,v}^{0}[\chi
_{v}[(\varepsilon ^{2}\Delta +\lambda )(\chi R_{\lambda
}^{0}f_{1})-f_{1}]].  \label{para}
\end{equation}
Here $L_{\tau ,d}^{2}$ is a subspace of $L^{2}(\Omega _{\varepsilon })$
which consists of functions supported on $\cup \Delta _{j}$. Now we are
going to define and study, successively, each of the terms in the formula
above. In particular, we need to show that
\begin{equation}
-(\varepsilon ^{2}\Delta +\lambda )P_{\lambda }f=f+Q_{\lambda }f,\text{ \ \
\ }Q_{\lambda }:L_{\tau ,d}^{2}\rightarrow L_{\tau ,d}^{2},\text{ \ \ \ }%
||Q_{\lambda }||\leq Ce^{\frac{-\rho }{\varepsilon }}.  \label{s}
\end{equation}

Operator $\widehat{G}_{\lambda }$ is an integral operator with the kernel $%
G_{\lambda }(x,z;\varepsilon ),$ $x,z\in \Omega _{\varepsilon },$\ which is
defined as follows. We split $\Omega _{\varepsilon }$ onto domains $\Omega
_{v,\varepsilon }$ by cutting all the finite channels $C_{j,\varepsilon }$
using the cross-sections $t=t_{j}.$ Let $z\in \Delta _{j_{0}}.$ Then
we choose $t_{j_{0}}\ $\ to be equal to the coordinate $t=t(z)$ of the point
$z.$ Other cross-sections are chosen with the only condition that $\tau
<t_{j}<l_{j}-\tau ,$ i.e., the cross-section $t=t_{j}$ is strictly inside of $%
\Delta _{j}$. Let $\Omega _{v,\varepsilon }^{\prime }$ be the spider domain
which we get by extending all the finite channels of $\Omega _{v,\varepsilon
}$ to infinity. Let $\Psi _{p,v}^{(\varepsilon )}$ be the scattering
solutions of the problem in the spider domain $\Omega _{v,\varepsilon
}^{\prime }.$ The small $\varepsilon $ asymptotics of these solutions is
given by Theorems \ref{tsp} and \ref{t6}. We introduce the following functions $%
\widetilde{\Psi }_{p,v}^{(\varepsilon )}$ by modifying the remainder terms
in these asymptotics:
\begin{equation}
\widetilde{\Psi }_{p,v}^{(\varepsilon )}=\psi _{p}\varphi _{0}(y/\varepsilon
)+\chi _{v}r_{p}^{(\varepsilon )},  \label{09}
\end{equation}
where $\chi _{v}\in C^{\infty }(\Omega _{\varepsilon }),$ $\chi _{v}=1$ on $%
\tau $-neighborhood of the junction, $\chi _{v}=0$ outside of $\Omega
_{v,\varepsilon }.$ Then we define $G_{\lambda }$ by the formula
\begin{equation}
G_{\lambda }(x,z;\varepsilon )=\Sigma _{p}\frac{a_{p,v}(\lambda ,\varepsilon
,\xi )}{D(\lambda ,\varepsilon )}\widetilde{\Psi }_{p,v}^{(\varepsilon )},%
\text{ \ \ \ \ }x\in \Omega _{v,\varepsilon },  \label{010}
\end{equation}
where $a_{p,v},D$ are defined in Lemmas \ref{lgr}, \ref{tr}, $\xi $ is the
point on the graph $\Gamma $ which corresponds $z\in \Delta _{j_{0}}$, i.e.,
the point on the edge $\Gamma _{j_{0}}$ where $t=t_{j_{0}}.$ Since function $%
\Psi _{p,v}^{(\varepsilon )}$ satisfies the equation $(\varepsilon
^{2}\Delta +\lambda )u=0$ on $\Omega _{v,\varepsilon },$ from Theorems \ref
{tsp} and \ref{t6} it follows that
\begin{equation*}
-(\varepsilon ^{2}\Delta +\lambda )\widetilde{\Psi }_{p,v}^{(\varepsilon
)}=O(\delta ^{-1}e^{-\frac{\alpha \tau }{\varepsilon }}),\text{ \ \ }%
\varepsilon \rightarrow 0,\text{ \ }-\infty <\lambda <\lambda ^{\prime }%
\text{ ,\ \ }x\in \Omega _{v,\varepsilon },
\end{equation*}
where $\alpha =\lambda _{1}-\lambda ^{\prime }.$ We choose $\nu <\frac{%
\tau }{4}.$ Then $\delta >e^{-\frac{\alpha \tau }{4\varepsilon }}$ for $%
\lambda \in (-\infty ,\lambda ^{\prime })\backslash $ $\Lambda ^{\nu }$, and
\begin{equation*}
-(\varepsilon ^{2}\Delta +\lambda )\widetilde{\Psi }_{p,v}^{(\varepsilon
)}=O(e^{-\frac{3\alpha \tau }{4\varepsilon }}),\text{ \ \ }\varepsilon
\rightarrow 0,\text{ \ }\lambda \in (-\infty ,\lambda ^{\prime })\backslash
\text{ }\Lambda ^{\nu }\text{,\ \ }x\in \Omega _{v,\varepsilon }.
\end{equation*}
Since coefficients $a_{p,v}$ are bounded, Lemma \ref{ld} with $\sigma =e^{-%
\frac{\alpha \tau }{4M\varepsilon }}$ implies that
\begin{equation}
-(\varepsilon ^{2}\Delta +\lambda )G_{\lambda }=O(e^{-\frac{\alpha \tau }{%
2\varepsilon }}),\text{ \ \ }\varepsilon \rightarrow 0,\text{ \ }\lambda \in
(-\infty ,\lambda ^{\prime })\backslash \text{ }\Lambda ^{\nu }\text{,\ \ }%
x\in \Omega _{v,\varepsilon }.  \label{101}
\end{equation}

Relations (\ref{101}) are valid on each domain $\Omega _{v,\varepsilon }.$
Now we are going to combine them and evaluate $(\varepsilon ^{2}\Delta
+\lambda )G_{\lambda }$ for all $x\in \Omega _{\varepsilon }.$ From (\ref{09}%
), (\ref{010}) and Lemma \ref{lgr} it follows that the function
\begin{equation*}
G_{\lambda }-g_{\lambda }(\gamma ,\xi ;\varepsilon )\varphi _{0}(\frac{y}{%
\varepsilon })
\end{equation*}
is infinitely smooth in the channels of $\Omega _{\varepsilon }.$ Here $%
\gamma $ is the point on $\Gamma $ which corresponds to $x\in \Omega
_{\varepsilon }.$ Then from (\ref{101}) it follows that
\begin{equation}
-(\varepsilon ^{2}\Delta +\lambda )G_{\lambda }=\delta _{\xi }(\gamma
)\varphi _{0}(\frac{y}{\varepsilon })+O(e^{-\frac{\alpha \tau }{2\varepsilon
}}),\text{ \ \ }\varepsilon \rightarrow 0,\text{\ }\lambda \in (-\infty
,\lambda ^{\prime })\backslash \text{ }\Lambda ^{\nu }\text{,\ \ }x\in
\Omega _{\varepsilon }.  \label{102}
\end{equation}
As it is easy to see, the remainder in (\ref{102}) is zero in the
region where $\nabla \chi _{v}\neq 0,$ i.e., support of the remainder belongs
to $\cup \Delta _{j}.$

\bigskip Now let us study the second and third terms in the left hand side
of (\ref{para}). Obviously,
\begin{equation}
-(\varepsilon ^{2}\Delta +\lambda )(\chi R_{\lambda }^{0}f_{1})=\chi
f_{1}+h=f_{1}+h,\text{ \ \ }h=-2\varepsilon ^{2}\nabla \chi \cdot \nabla
R_{\lambda }^{0}f_{1}-\varepsilon ^{2}(\Delta \chi )R_{\lambda }^{0}f_{1}.
\label{xx}
\end{equation}
Here we used the fact that $\chi =1$ on the support of $f_{1}.$ Since $f_{1}$
is orthogonal to $\varphi _{0}(\frac{y}{\varepsilon }),$ function $%
R_{\lambda }^{0}f_{1}$ and all its derivatives decay exponentially in each
channel $C_{j,\varepsilon }$ as $\frac{r}{\varepsilon }\rightarrow \infty $
where $r$\ \ is the distance from $\Delta _{j.}$ Hence,
\begin{equation}
h=O(e^{-\frac{\alpha (\tau -\varepsilon )}{\varepsilon }})=O(e^{-\frac{%
\alpha \tau }{\varepsilon }}),\text{ \ \ \ }\varepsilon \rightarrow 0,\text{
\ }\lambda \in (-\infty ,\lambda ^{\prime }).  \label{hv}
\end{equation}

The remainder terms will be parts of the operator $Q_{\lambda },$ and we
need the kernel of this operator to be supported on $\cup \Delta _{j}.$
Unfortunately, $h$ is supported on $\varepsilon $- neighborhoods of the
junctions. The last term in (\ref{para}) is designed to correct this.
Since $h$ is supported on the region where $\nabla \chi \neq 0,
$ function $h$ can be represented as the sum $h=\Sigma _{v}h_{v}$, where $%
h_{v}=\chi _{v}h$ has estimate (\ref{hv}) and is supported on the $%
\varepsilon $-neighborhood of the junction $J_{v,\varepsilon }$ which
corresponds to the vertex $v.$ Consider $\widetilde{h}=\Sigma _{v}\chi
_{v}R_{\lambda ,v}^{0}[\chi _{v}h]$ which is defined as follows. We apply
the resolvent $R_{\lambda ,v}^{0}$ of the problem in the spider domain $%
\Omega _{v,\varepsilon }^{\prime }$ to $h_{v},$ multiply the result by $\chi
_{v}$ and extend the product by zero on $\Omega _{\varepsilon }\backslash
\Omega _{v,\varepsilon }.$

From (\ref{hv}) and Theorem \ref{tsp} it follows that
\begin{equation}
|R_{\lambda ,v}^{0}h_{v}|\leq C\delta ^{-1}e^{-\frac{\alpha \tau }{%
\varepsilon }}\leq Ce^{-\frac{\alpha \tau }{2\varepsilon }},\text{ \ \ \ }%
\varepsilon \rightarrow 0,\text{\ }\lambda \in (-\infty ,\lambda ^{\prime
})\backslash \text{ }\Lambda ^{\nu },  \label{v1}
\end{equation}
if we choose $\nu <\frac{\tau }{2},$ so that $\delta >e^{-\frac{\alpha \tau
}{2\varepsilon }}.$ From standard a priory estimates for the solutions of
homogeneous equation $(\varepsilon ^{2}\Delta +\lambda )u=0$ it follows that
estimate (\ref{res1}) is valid also for all derivatives of $R_{\lambda }f,$
since this function satisfies the homogeneous equation outside of $%
2\varepsilon $-neighborhood of the junction. Then (\ref{v1}) holds for the
derivatives of $R_{\lambda ,v}^{0}h_{v}.$ This allows us to obtain, similarly
to (\ref{xx}), that
\begin{equation}
-(\varepsilon ^{2}\Delta +\lambda )\widetilde{h}=h+h_{1},\text{ \ \ }%
h_{1}=O(e^{-\frac{\alpha \tau }{2\varepsilon }}),\text{ \ \ \ }\varepsilon
\rightarrow 0,\text{ \ }\lambda \in (-\infty ,\lambda ^{\prime })\backslash
\text{ }\Lambda ^{\nu },  \label{v2}
\end{equation}
where $h_{1}$ is supported on the closure of the set $\nabla \chi _{v}\neq 0.
$ This set belongs to $\cup \Delta _{j}.$ Finally, from (\ref{102}), (\ref{xx}%
), (\ref{v2}) it follows that
\begin{equation}
-(\varepsilon ^{2}\Delta +\lambda )P_{\lambda }f=f+g,\text{ \ \ }g=O(e^{-%
\frac{\rho }{\varepsilon }}),\text{ \ \ \ }\varepsilon \rightarrow 0,\text{
\ }\lambda \in (-\infty ,\lambda ^{\prime })\backslash \text{ }\Lambda ^{\nu
},  \label{gg}
\end{equation}
and $g$ is supported on $\cup \Delta _{j}.$ One can easily check that $g$
depends linearly on $f.$ Besides, one can specify the dependance on the norm of $f$
in estimates of all the remainders above. This will lead to (\ref{s}) instead
of (\ref{gg}). In fact, (\ref{s}) is valid when $Q_{\lambda }$ is considered
as an operator in $L^{2}$ or as an operator in the space of continuous functions
on $\cup \Delta _{j}.$

We are going to construct now the solution $u$ of problem (\ref{a1}) with $%
f\in $ $L_{\tau ,d}^{2}$. We look for $u$ in the form $u=P_{\lambda }g$ with
unknown $g\in $ $L_{\tau ,d}^{2}.$ Obviously, $u$ satisfies the boundary
conditions and appropriate conditions at infinity. Equation (\ref{a1}) in $%
\Omega _{\varepsilon }$ leads to $g+Q_{\lambda }g=f.$ Since the norm of
operator $Q_{\lambda }$ is exponentially small, function $g$ exists, is unique
and $g=f+q,$ $||q||\leq Ce^{-\frac{\rho }{\varepsilon }}||f||,$ i.e.,
\begin{equation*}
u=P_{\lambda }(f+q),\text{ \ \ }||q||_{L_{\tau ,d}^{2}}\leq Ce^{-\frac{\rho
}{\varepsilon }}||f||_{L_{\tau ,d}^{2}},\text{ \ \ }\varepsilon \rightarrow
0,\text{ \ }\lambda \in (-\infty ,\lambda ^{\prime })\backslash \text{ }%
\Lambda ^{\nu }.
\end{equation*}
This justifies (\ref{lll}) and (\ref{rem}). The first statement of Theorem
\ref{tR1} follows from here. Namely, assume that an eigenvalue $\mu =\mu
_{j,\varepsilon }$ of the operator $H_{\varepsilon }$ belongs to $(-\infty
,\lambda ^{\prime })\backslash $ $\Lambda ^{\nu }.$ Then the truncated resolvent
$R_{\lambda }$ (see (\ref{res})) has a pole there (see Theorem \ref{t1}).
The residue of this pole is the orthogonal projection on the eigenspace of $%
H_{\varepsilon }$. The pole of $R_{\lambda }f$ may disappear only if $f$ is
orthogonal to the eigenspace which corresponds to the eigenvalue $\lambda
=\mu .$ Non-trivial solutions of the equation $(\Delta +\lambda )u=0$ in $%
\Omega _{\varepsilon }$ can not be equal to zero in a subdomain of $\Omega
_{\varepsilon }.$ Thus, there is a function $f\in $ $L_{\tau ,d}^{2}$ which
is not orthogonal to the eigenspace, and $R_{\lambda }f$ must have a pole at
$\lambda =\mu .$ This contradicts (\ref{lll}) and (\ref{rem}). \qed

The following statement can be easily proved using Theorem \ref{tR1} and
reduction (\ref{fu}) of the scattering problem to problem (\ref{a1}), (\ref
{a2}).

\begin{theorem}
For any interval $[\alpha ,\lambda ^{\prime }),$there exist $\rho ,\nu >0$
such that scattering solutions $\Psi _{p,\varepsilon }(x)$ of the problem in
$\Omega \varepsilon $ have the following asymptotic behavior on the channels
of $\Omega \varepsilon $ as $\varepsilon \rightarrow 0$%
\begin{equation*}
\Psi (x)=\psi _{p,\varepsilon }(\gamma )\varphi _{0}(\frac{y}{\varepsilon }%
)+r_{p}^{(\varepsilon )}(x),
\end{equation*}
where $\psi _{p}(\gamma )=\psi _{p}^{(\varepsilon )}(\gamma )$ are the
scattering solutions  of the problem on the graph $\Gamma $ and
\begin{equation*}
|r_{p}^{(\varepsilon )}(x)|\leq Ce^{\frac{-\rho d(\gamma )}{\varepsilon }},%
\text{ \ \ \ \ \ \ }\lambda \in \lbrack \alpha ,\lambda ^{\prime
})\backslash \text{ }\Lambda ^{\nu }.
\end{equation*}
Here $\gamma =\gamma (x)$ is the point on $\Gamma $ which is defined by the
cross-section of the channel through the point $x,$ and $d(\gamma )$ is the
distance between $\gamma $ and the closest vertex of the graph.
\end{theorem}
\section{Eigenvalues near the threshold }

In some cases, in particular when the parabolic problem is studied, the
lower part of the spectrum of the operator $H_{\varepsilon }$ is of a
particular importance. Theorem \ref{tR1} provides a full description of the
location of the eigenvalues. One may have a finite number of eigenvalues
below $\lambda _{0}.$ They are determined by the junctions and situated in
an exponentially small neighborhood of $\Lambda ^{0}$ (set of eigenvalues of
the corresponding spider domains). The eigenvalues between $\lambda _{0}$
and $\lambda _{1}$  are situated in an exponentially small neighborhood of $%
\Lambda (\varepsilon )$ which is the set of eigenvalues of the one
dimensional problem (\ref{spp})-(\ref{bes}) on the limiting graph (see
Theorem \ref{tr}).

We will assume that $\Omega _{\varepsilon }$ has at least one bounded
channel (for example, $\Omega _{\varepsilon }$ is bounded). The opposite case
is studied in Theorem \ref{tsp}. We also assume that $\lambda =\lambda
_{0}+O(\varepsilon ^{2}).$ Then the eigenvalues of the problem (\ref{spp})-(%
\ref{bes}) will depend on the form of the GC (\ref{gc}) at $\lambda =\lambda
_{0}.$ Let put $\lambda =\lambda _{0}+\mu \varepsilon ^{2}$ in (\ref{spp})-(%
\ref{bes}). Then this problem takes the form
\begin{equation}
-\frac{d^{2}}{dt^{2}}\varsigma =\mu \varsigma \text{ \ \ \ \ on }\Gamma ,
\label{lp1}
\end{equation}
\begin{equation}
B\varsigma =0\text{ \ \ at }v\in V_{1},  \label{lp2}
\end{equation}
\begin{equation}
i[I_{v}+T_{v}(\lambda _{0}+\mu \varepsilon ^{2})]\frac{d}{dt}\varsigma
^{(v)}(t)-\mu \lbrack I_{v}-T_{v}(\lambda _{0}+\mu \varepsilon
^{2})]\varsigma ^{(v)}(t)=0,\text{ \ \ \ }t=0,\text{ \ \ \ }v\in V_{2},
\label{lp3}
\end{equation}

\begin{equation}
\varsigma =a_{j}e^{i\mu t},\text{ \ }\gamma \in \Gamma _{j},\text{ \ \ }%
1\leq j\leq m,~~t>>1.  \label{lp4}
\end{equation}
The last condition is not needed if  $\Omega _{\varepsilon }$ is bounded ($%
m=0$).

Since matrix $T_{v}(\lambda _{0})$ is orthogonal and its eigenvalues are $%
\pm 1,$ the GC\ (\ref{lp3}) with $\varepsilon =0$ has the form
\begin{equation}
P\varsigma ^{(v)}(0)=0,\text{ \ \ }P^{\perp }\frac{d}{dt}\varsigma
^{(v)}(0)=0,\text{ \ \ \ }v\in V_{2},  \label{lp31}
\end{equation}
where $P,P^{\perp }$ are projections onto eigenspaces of matrix $%
T_{v}(\lambda _{0})$ with the eigenvalues $\mp 1,$ respectively. Let $k$ be
the dimension of the operator $P,$ and $d-k$ be the dimension of the
operator $P^{\perp },$ where $d=d(v)$ is the size of the vector $\varsigma
^{(v)}.$ Then (\ref{lp31}) imposes $k$ Dirichlet conditions and $d-k$
Neumann conditions on the components of vector $\varsigma ^{(v)}$ written in the
eigenbasis of the matrix $T_{v}(\lambda _{0}).$ \ Note that the standard
Kirchhoff conditions ($\varsigma $ is continues on $\Gamma ,$ a linear
combination of derivatives is zero at each vertex) has the same nature, and $%
k=d-1$ in this case.

Problem (\ref{lp1})-(\ref{lp4}) with $\varepsilon =0$ has a discrete
spectrum $\{\mu _{j}\}$, $j\geq 1,$ and the same problem with $\varepsilon >0
$ is its analytic perturbation. Thus, the following statement is valid.

\begin{theorem}
\label{tan}If eigenvalues $\{\mu _{j}\}$ are simple, then eigenvalues $\{\mu
_{j}(\varepsilon )\}$ of problem (\ref{lp1})-(\ref{lp4}) are analytic in $%
\varepsilon :$%
\begin{equation}
\mu _{j}\text{(}\varepsilon \text{)}=\sum_{n\geq 0}\mu _{j,n}\varepsilon
^{n},\text{ \ \ }\mu _{j,0}=\mu _{j}.  \label{expq}
\end{equation}
\end{theorem}
\noindent \textbf{Remarks} 1. This statement implies that eigenvalues $\lambda \in
\Lambda (\varepsilon )$ in $O(\varepsilon ^{2})$-neighborhood of $\lambda
_{0}$ have the form
\begin{equation}
\lambda =\lambda _{j}(\varepsilon )=\lambda _{0}+\varepsilon ^{2}\sum_{n\geq
0}\mu _{j,n}\varepsilon ^{n}.  \label{expp}
\end{equation}
2. The assumption on simplicity of $\mu _{j}$ often can be omitted. For example (\ref
{expq}),(\ref{expp}) remain valid without this assumption if $k=d$ (the limiting problem is the
Dirichlet problem). In the latter case one may have multiple eigenvalues (for
example, when the graph has edges of multiple lengths), but the problem with
$\varepsilon =0$ is split into separate problems on individual edges.

Theorem \ref{tan} makes it important to specify the value of $k$ in the
condition (\ref{lp31}). This value depends essentially on the type of the
boundary conditions at $\partial \Omega _{\varepsilon }$ and on whether $%
\lambda =\lambda _{0}$ is a pole of the truncated resolvent (\ref{b2}) or
not.

\begin{definition}
A ground state of the operator $H_{\varepsilon }$ in a domain $\Omega
_{\varepsilon }$ at $\lambda =\lambda _{0}$ is the function $\psi _{0}=\psi
_{0}\left( x\right) $, which is bounded, strictly positive inside $\Omega
_{\varepsilon }$, satisfies the equation $\left( -\Delta -\lambda
_{0}\right) \psi _{0}=0$ in $\Omega _{\varepsilon }$, and the boundary
condition on $\partial \Omega _{\varepsilon },$ and has the following
asymptotic behavior at infinity
\begin{equation}
\psi _{0}\left( x\right) =\varphi _{0}\left( \frac{y}{\varepsilon }\right)
[\rho _{j}+o\left( 1\right) ],\text{ \ }x\in C_{j},\text{ \ }|x|\rightarrow
+\infty ,  \label{1b1}
\end{equation}
where $\rho _{j}>0$ and $\varphi _{0}$ is the ground state of the operator
in the cross-sections of the channels$.$
\end{definition}

Obviously, if the Neumann boundary condition is imposed on $\partial \Omega
_{\varepsilon },$ then $\lambda _{0}=0$, and the ground state at $\lambda =0$
exists and equal to a constant. It was shown in \cite{MV1}, \cite{MV2} that
the ground state at $\lambda =\lambda _{0}$ does not exist for generic
domains $\Omega _{\varepsilon }$ in the case of other boundary conditions on
$\partial \Omega _{\varepsilon }.$ In particular, it does not exist if there
are eigenvalues of $H_{\varepsilon }$ below $\lambda _{0}$ or if the
truncated resolvent does not have a pole at $\lambda =\lambda _{0}.$ The following result was proved in
 \cite{MV1}, \cite{MV2}.
\begin{theorem}
(1) The ground state at $\lambda =\lambda
_{0}$ implies $k=d-1.$ Thus the eigenvalues $\mu _{j}(\varepsilon ),$ $
\varepsilon \rightarrow 0,$\ converge to the eigenvalues of the Kirchhoff
problem in the case of Neumann condition on $\partial \Omega _{\varepsilon }$
($\Omega _{\varepsilon }$ is arbitrary) and in the case of other boundary
conditions on $\partial \Omega _{\varepsilon }$ for special, non-generic $%
\Omega _{\varepsilon }.$

(2) If Dirichlet or Robin condition is imposed on $%
\partial \Omega _{\varepsilon }$ and the truncated resolvent does not have a
pole at $\lambda =\lambda _{0}$ (this is a generic condition on $\Omega
_{\varepsilon }$), then $k=d$ and $\mu _{j}(\varepsilon ),$ $\varepsilon
\rightarrow 0,$\ converge to the eigenvalues of the Dirichlet problem.
\end{theorem}
Other possible (non-generic) GC  at $\lambda = \lambda_0$ are given by (\ref{lp31}).

\end{document}